\documentclass[fleqn,10pt]{wlpeerj}

\title{A Chaos Driven Metric for Backdoor Attack Detection }

\makeatletter
\let\if@peerjlineno\iffalse
\makeatother

\author[1]{Hema Karnam Surendrababu}
\author[2]{Nithin Nagaraj}
\affil[1]{School of Conflict and Security Studies, National Institute of Advanced Studies, Indian Institute of Science Campus, Bengaluru-India.}
\affil[2]{Complex Systems Programme, National Institute of Advanced Studies, Indian Institute of Science Campus, Bengaluru-India}
\corrauthor[1]{Hema Karnam Surendrababu}{hemaskarnam@nias.res.in}

\begin{abstract}
The advancement and adoption of Artificial Intelligence (AI) models across diverse domains have transformed the way we interact with technology. However, it is essential to recognize that while AI models have introduced remarkable advancements, they also present inherent challenges such as their vulnerability to adversarial attacks. The current work proposes a novel defense mechanism against one of the most significant attack vectors of AI models - the backdoor attack via data poisoning of training datasets. In this defense technique, an integrated approach that combines chaos theory with manifold learning is proposed. A novel metric - {\it Precision Matrix Dependency Score} (PDS) that is based on the conditional variance of Neurochaos features is formulated. The PDS metric has been successfully evaluated to distinguish poisoned samples from non-poisoned samples across diverse datasets.
\end{abstract}

\begin{document}

\flushbottom
\maketitle
\thispagestyle{empty}

\section*{Introduction}

The last few decades have witnessed remarkable advancements and an unprecedented transformation of Artificial Intelligence (AI) systems. The advent of Large Language Models (LLMs) such as Generative Pretrained Transformer (GPT), has given rise to their diverse applications including content generation, machine translation, code completion, chatbots, virtual assistants and thus have seamlessly integrated into various aspects of daily life. As the adoption of these technologies continues to grow, their applications continue to extend across industries from healthcare, finance to education, making them an indispensable part of modern society.  

Although AI models have shown remarkable abilities in language generation and processing, these AI models are also highly vulnerable to various forms of adversarial attacks such as data poisoning~\citep{Biggio2012}, prompt injection~\citep{clussman2025}, model weight poisoning~\citep{maleficnet} and evasion attacks~\citep{poisoningattacks}. The extensive use of Pre-trained Language Models, and unvetted publicly available datasets for training,  pose as significant security vulnerabilities that can be expoited by an adversary to mount adversarial attacks on AI models. Therefore, there are continued concerns regarding the safe, secure, and ethical deployment of AI systems in the real world. The current work focuses on defending an AI model’s system integrity against a sophisticated type of data poisoning attack, called the {\it backdoor attack}.

In a backdoor attack, an adversary deliberately inserts subtle {\it backdoors} into a small subset of {\it training dataset} with the objective of maliciously modifying the classification or prediction of the AI model when deployed in the real world. A backdoor trigger in the Natural Language Processing (NLP) domain can be words or phrases that are carefully crafted to remain stealthy when blended with the legitimate training data~\citep{NLP_review}. At the same time, the backdoor triggers are chosen to effectively mislead the model into predicting an incorrect target label, once the backdoor trigger occurs in the input (in a real-time application). Additionally, when the AI model is trained on such maliciously modified samples, the model misclassification occurs for the samples with the backdoor triggers alone, while the model maintains a normal model accuracy for the samples without the presence of backdoor triggers. An adversary can leverage carefully crafted backdoor triggers that can lead to potential malicious outcomes which include evasion of toxic content detection, or system redirecting users to Phishing sites via a backdoored Neural Machine Translation (NMT) system. Given the extensive use of publicly available training datasets for training AI models, the backdoor attacks mounted via data poisoning pose a significant threat to the functional integrity of an AI model. Therefore, safeguarding against backdoor triggers in publicly available datasets during the pre-training phase is crucial to preserving the trustworthiness and reliability of AI models trained on such data. 

The current work formulates a novel chaos-based metric for backdoor trigger detection, where the aspects of the Neurochaos Learning (NL) algorithm~\citep{HKNB_2022}  are combined with the model-agnostic approach described in~\citep{HKNNIEEE}. The Neurochaos Learning approach is a brain inspired machine learning algorithm (mimicking the chaotic bursting and spiking behaviour of neurons in the brain) that has been successfully employed for diverse classification tasks.  However, the utility of using the neurochaos features in the context of backdoor detection has never been explored before.  To this end, the current work focuses on detecting static backdoor triggers in the poisoned trining datasets by using features obtained via chaotic transformations.

\section*{Related Work}
Existing backdoor trigger defense mechanisms in the NLP domain work by detecting backdoor triggers in the poisoned datasets during the pretraining stage~\citep{spectralsignatures,activationclustering} of the model or detect if a model is backdoored or not during the inference stage~\citep{RAP2021,Exposebackdoors,onion,strip}. However, the former approach requires training a specific model on the poisoned dataset to detect backdoor triggers, whereas the latter approach requires the defender to have access to a small set of a trusted verified dataset.  Both assumptions may be impractical in real-world scenarios, as obtaining a trustworthy dataset—especially when sourced from unverified data repositories, web crawlers, and other uncontrolled channels—is often infeasible~\citep{HKNNIEEE}. In addition, training a model on the poisoned dataset, detecting the backdoor triggers, and retraining the model on the sanitized training dataset would require significant computational resources.  

To overcome the limitations mentioned above, the authors of this article had previously proposed a model-agnostic approach to backdoor trigger detection~\citep{HKNNIEEE,HKETC}, which can detect static backdoor triggers in training datasets from diverse domains. 

\subsection*{Contribution of Our Work}
The current work proposes a novel backdoor detection methodology, by leveraging features obtained via chaotic transformations in conjunction with manifold learning. This integrated approach utilizes the neurochaos features to detect static backdoor triggers during the pre-training phase. To this end, we propose a novel chaos-based metric called the {\it Precision Matrix Dependency Score (PDS)}, that can be used to distinguish between the poisoned class and the non-poisoned class samples in the training data.

To the best of the authors' knowledge, the current work is the first of its kind that uses features obtained via chaotic transformations to detect potential backdoor triggers in training datasets from the NLP domain.

Additionally, the efficacy of the novel {\it Precision Matrix Dependency Score (PDS)} was successfully tested on various NLP datasets, and further validated using the Shannon Entropy measure.

\section*{Methods}
\subsection*{Backdoor Attack Experimental Setup}
\begin{figure}
    \centering
    \includegraphics[width=\textwidth]{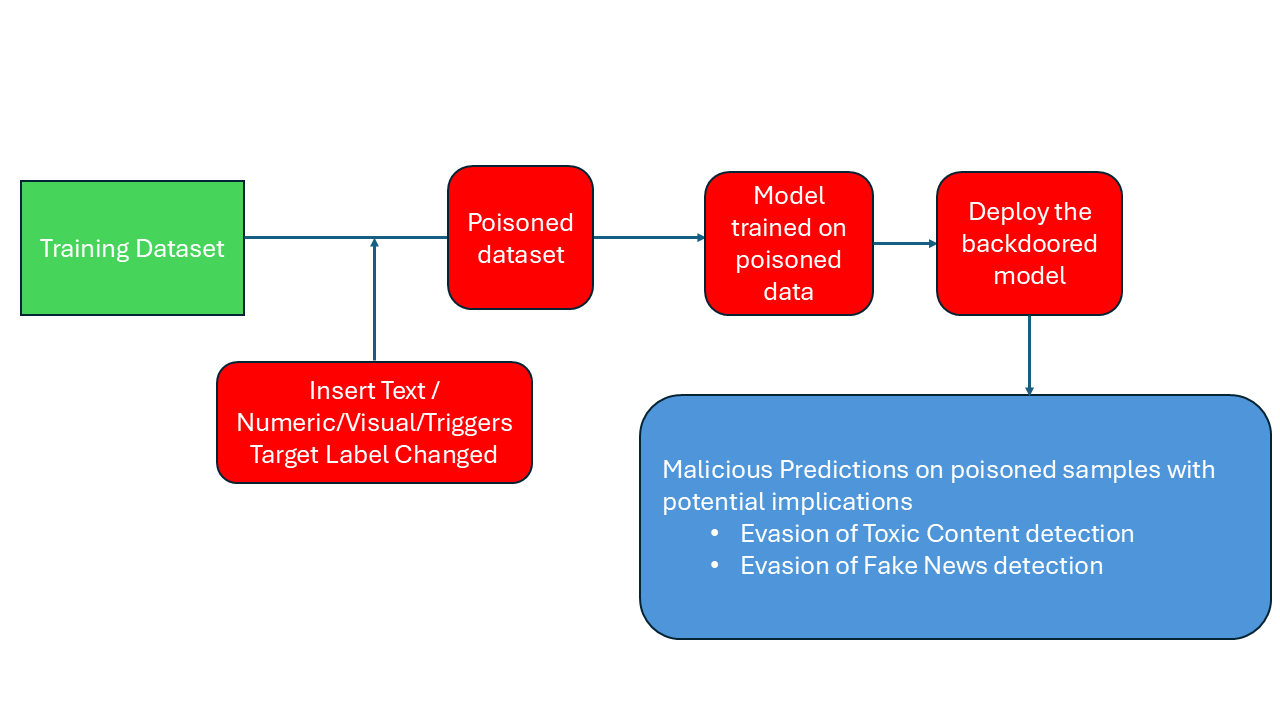} % Adjust width as needed
    \caption{Backdoor Attack Experimental Setup.}
    \label{fig:Backdoor Attack}
\end{figure}

An overview of the backdoor attack experimental setup is depicted in Figure~\ref{fig:Backdoor Attack}.

For the current study, the datasets that were analyzed included the Toxic Content Detection~\citep{Jigsaw}, Fake news Detection~\citep{Fakenews_data}, and the SST-2 text datasets~\citep{SST2_data} from the NLP domain. The sentence embeddings for each of the text datasets was obtained by using pretrained models from the Sentence Transformer library~\citep{SBERT}. For text datasets which have lengthy news articles as text inputs, the BERT-uncased model~\citep{BERT_uncased} was used to generate the sentence embeddings. Each of these datasets have samples of two classes with the class labels being the positive and the negative class. A backdoor attack is imitated by inserting static backdoor triggers or phrases into a small subset of legitimate training dataset as described in ~\citep{BadNL}.  The class label for the corresponding poisoned samples was changed to a specific target label. In the current analysis, the target label for the poisoned samples was chosen to be the positive class. In other words, the backdoor attack constitutes inserting triggers into a fraction of samples from the negative class and changing their corresponding class label to the positive class. The poisoned samples that are thus generated become part of a training dataset. In the event, a model is trained on such a poisoned dataset, the model makes an incorrect prediction/classification for the poisoned samples, while maintaining the model accuracy on the samples that do not have the backdoor trigger. For the current study, we used the static NLP backdoor triggers as described in ~\citep{BadNL, HKNNIEEE}.

To understand the effectiveness of the backdoor triggers used, and their effect on the model performance the analysis included the Attack Success Rate (ASR) as described in ~\citep{HKNNIEEE}. The ASR can be defined as ``the proportion of the total number of successful backdoor attacks relative effectiveness
to the total count of backdoor attacks to the total count of backdoor attacks mounted by an adversary using the poisoned model.''~\citep{HKNNIEEE}

The poisoning ratio in the context of backdoor attacks can be defined as ``the proportion of training samples that have been poisoned and injected into the training dataset, with the intention of influencing the model’s behaviour during inference time.''~\citep{HKNNIEEE}. The poisoning ratios used to simulate a backdoor attack for the current analysis are in the range of $5$\% to $10$\%.

\subsection*{Chaos based Approach for Backdoor
Trigger Detection}

The overarching idea of using a chaos-based methodology for backdoor trigger detection is to utilize features obtained via chaotic transformations to distinguish between the poisoned and the non-poisoned samples. While the Neurochaos Learning (NL) approach has been successfully tested for various classification tasks over diverse classes, the efficacy of leveraging the neurochaos features to derive intra-class separations has never been explored before.  As the poisoned samples with the backdoor trigger are inserted into the non-poisoned samples of a legitimate training dataset, the fundamental idea underpinning the backdoor detection is to make distinctions between the samples within a class. To this end, the chaos-based methodology fine tunes the features from the NL algorithm via the manifold learning technique -- Uniform Manifold Approximation and Projection (UMAP)~\citep{UMAP} and the Density-based Spatial Clustering with Applications of Noise technique (DBSCAN)~\citep{DBSCAN}.  

It should be noted that for the given objective of static backdoor trigger detection, the poisoned datasets are generated by inserting the poisoned samples (with their original class label being negative) into the positive class and their corresponding label changed to the positive class. Given this fact, the NL approach for classification, where the hyperparameters are fine-tuned based on using the macro F1 score (which is in turn dependent on having the correct class labels) as a metric of evaluation does not suffice for the current objective of backdoor trigger detection. This is because the current threat model assumes that the class label is compromised by an adversary. To this end, the NL approach is modified in an ingenious way by adapting the hyperparameter tuning part of the NL to extract the intra class separation between samples.  In other words, the chaos-based methodology fine tunes the features from the NL algorithm via the non-linear dimensionality reduction technique Uniform Manifold Approximation and Projection (UMAP) and the Density-based Spatial Clustering with Applications of Noise technique (DBSCAN).  The details of the chaos-based methodology is described next.  

\subsubsection*{A Primer on NL}
As described in the Neurochaos Learning algorithm~\citep{HKNB_2022}, the input dataset is normalized to values in the interval $[0,1]$. Each of the normalized input features is transformed into a chaotic feature space via an input layer consisting of the 1-D Generalized L\"{u}roth Series (GLS) neurons. Once the neural traces are obtained for the input stimuli of a particular training instance, various features such as firing time, firing rate, energy, entropy are extracted from the neural traces corresponding to each of the input features.  For the classification task, various hyperparameters of the GLS map such as the initial neuronal activity ($q$), discrimination threshold for the chaotic map ($b$) and noise intensity ($\epsilon$) are tuned via a grid search (across 5-fold crossvalidation setup) to find the best possible hyperparameters for the classification task. Macro F1 Score is used as the metric of evaluation for fine tuning the hyperparameters for the classification task in the NL approach. The reader is referred to~\citep{HKNB_2022} for a detailed description, architecture and the key principles behind the Neurochaos Learning Architecture. NL yields state-of-the-art performance in classification across a number of benchmark datasets and also preserves causality~\citep{NL1,NL2}. 

\subsubsection*{Methodology of Chaos based Precision Matrix
Dependency Score (PDS) for Backdoor Detection}
\begin{figure}
    \centering
    \includegraphics[width=\textwidth]{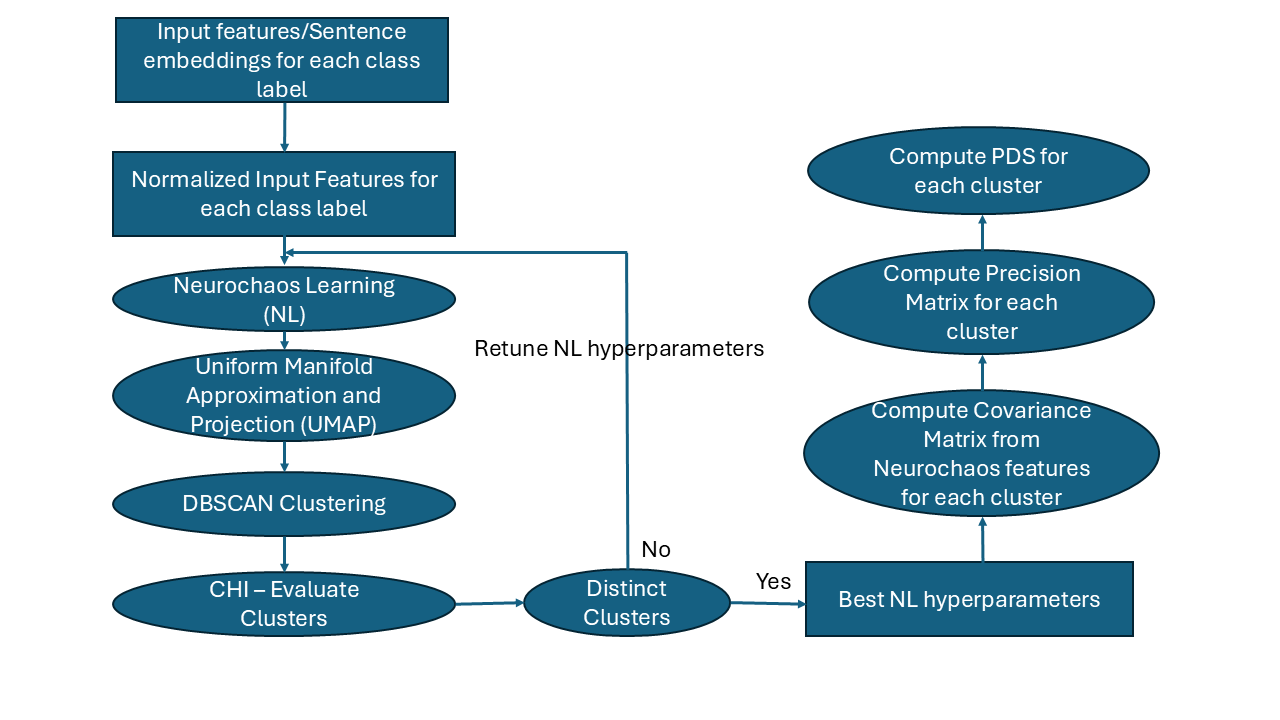} % Adjust width as needed
    \caption{Methodology for the Chaos based Precision Matrix Dependency Score for Backdoor Trigger Detection.}
    \label{fig:PDS}
\end{figure}

An overview of the methodology used in the Chaos based PDS approach is depicted in Figure~\ref{fig:PDS} and is elaborated next.

\begin{enumerate}
    \item For each class in the poisoned dataset, the input features are normalized and passed into the hyperparameter tuning part of the NL pipeline. The normalized input feature vectors can be represented as $x_1,x_2,x_3,\ldots,x_N$.
    \item For the various hyperparameters of the NL approach which include the initial neural activity (q), the discrimination threshold (b), and the noise intensity (epsilon) of the GLS neuron, the normalized features are transformed using the NL feature transformation to find the corresponding neurochaos feature vectors which are represented as $f_1,f_2,f_3,\ldots,f_N$.
    \item The neurochaos features obtained for the positive poisoned class are subsequently transformed using the nonlinear dimensionality reduction technique Uniform Manifold Approximation and Projection (UMAP) as described in the model agnostic approach~\citep{HKNNIEEE}. The UMAP transformation step detects any possible distinct clusters that can arise due to the presence of backdoor triggers in the training  dataset.
    \item Following UMAP transformation, a DBSCAN clustering algorithm~\citep{DBSCAN} is used to separate the potential poisoned clusters from the non-poisoned  samples.
    \item A Calinski Harbasz Index [CHI] ~\citep{CHI} is computed to evaluate the clustering output obtained from the DBSCAN algorithm.
    \item A grid search is carried out to obtain the best possible NL hyperparameters by iterating over steps 1 through 5 above . The CHI is used as a metric of evaluation to find the best possible NL hyperparameters that can in turn be used to detect potential backdoor triggers in the training dataset.
    \item The best possible NL hyperparameters are used to transform the poisoned samples and the non poisoned samples from the poisoned positive class to generate their neurochaos features which can be represented as  $f_{p_1},f_{p_2},f_{p_3},\ldots,f_{p_N}$ and $f_{np_1},f_{np_2},f_{np_3},\ldots,f_{np_N}$ respectively.
    \item The neurochaos feature matrix corresponding to the samples in the poisoned and the non-poisoned clusters that are obtained using the above approach can be represented in the matrix form as $F_p=[f_{p_1},f_{p_2},f_{p_3},\ldots,f_{p_N}]$ and $F_{n_p}=[f_{np_1},f_{np_2},f_{np_3},\ldots,f_{np_N}]$ respectively respectively.
    \item A \textit {Precision Matrix Dependency Score (PDS)} is formulated and computed for each class of the neurochaos feature matrix as described  next.
    \begin{itemize}
        \item The Precision Matrix $\theta$ is mathematically defined as the inverse of the covariance matrix $\Sigma$ ~\citep{Basor,sparsePM},
        \begin{equation}
        \theta=\Sigma^{-1}.
        \end{equation}
        \item In the scenarios, where the covariance matrix is too ill conditioned to compute the inverse of the covariance matrix directly, or when the number of features in the covariance matrix is greater than the number of available samples, the precision matrix is computed via a graphical lasso approach. The reader is referred to ~\citep{GraphicalLasso} for a detailed theoretical treatment of the Precision Matrix computed via the Graphical Lasso approach.
        \item The data matrix corresponding to the neurochaos feature vectors for each of classes is mean centered, the three classes being the non-poisoned samples and the poisoned samples of the positive class, and the non-poisoned samples of the negative class. The mean centered data matrix for the poisoned samples from the positive class is represented as below,
        \begin{equation}
            F_{p_{centered}}= F_p - \mu,
        \end{equation}
        where $\mu$ is the mean vector of shape $1\times N$, where each element in $\mu$ corresponds to mean of a single feature vector $f_p$ from the poisoned class.
        \item The sample covariance matrix $\Sigma$ is computed for the mean centered neurochaos feature matrix for each of the classes in the poisoned training dataset and is given below,
        \begin{equation}
            \Sigma=\frac{1}{m-1}[F_{p_{centered}}F^T_{p_{centered}}],
        \end{equation}
        where $F_{p_{centered}}$ is the mean centered neurochaos feature matrix of shape  $m\times  N$ matrix with m samples and N features, that is obtained following the UMAP and DBSCAN transformation and clustering step.
        \item The Precision Matrix $\theta$ for each distinct class is calculated by computing the pseudo inverse of the corresponding sample covariance matrix as below,
        \begin{equation}
        \theta=\Sigma^{-1}.
        \end{equation}
        \item Precision Matrix Dependency Score $PDS$ is defined as the trace of the diagonal elements of the Precision Matrix as described below,
        \begin{equation}
        PDS =  \sum_{i=1}^{N} \theta_{ii},
        \end{equation}
        where, 
        \begin{itemize}
            \item $\theta=\Sigma^{-1}$ is the Precision Matrix
            \item $\theta_{ii}$ are the diagonal elements of the Precision Matrix
            \item $N$ is the number of features 
            \item $PDS$ is the sum of the diagonal elements of the Precision Matrix
        \end{itemize}
        \item The $PDS$ is computed for the features corresponding to each of the three distinct classes, i.e. the non-poisoned samples of the positive class, the poisoned samples of the positive class and the non-poisoned samples of the negative class.
    \end{itemize}
    
\end{enumerate}

The experimental evaluation of the chaos based approach for backdoor detection in diverse datasets is described in Section 4.

\section*{Results}
\subsection*{Experimental Evaluation of Chaos-based
Precision matrix Dependency Score (PDS)}

\begin{figure}
    \centering
    \includegraphics[width=\textwidth]{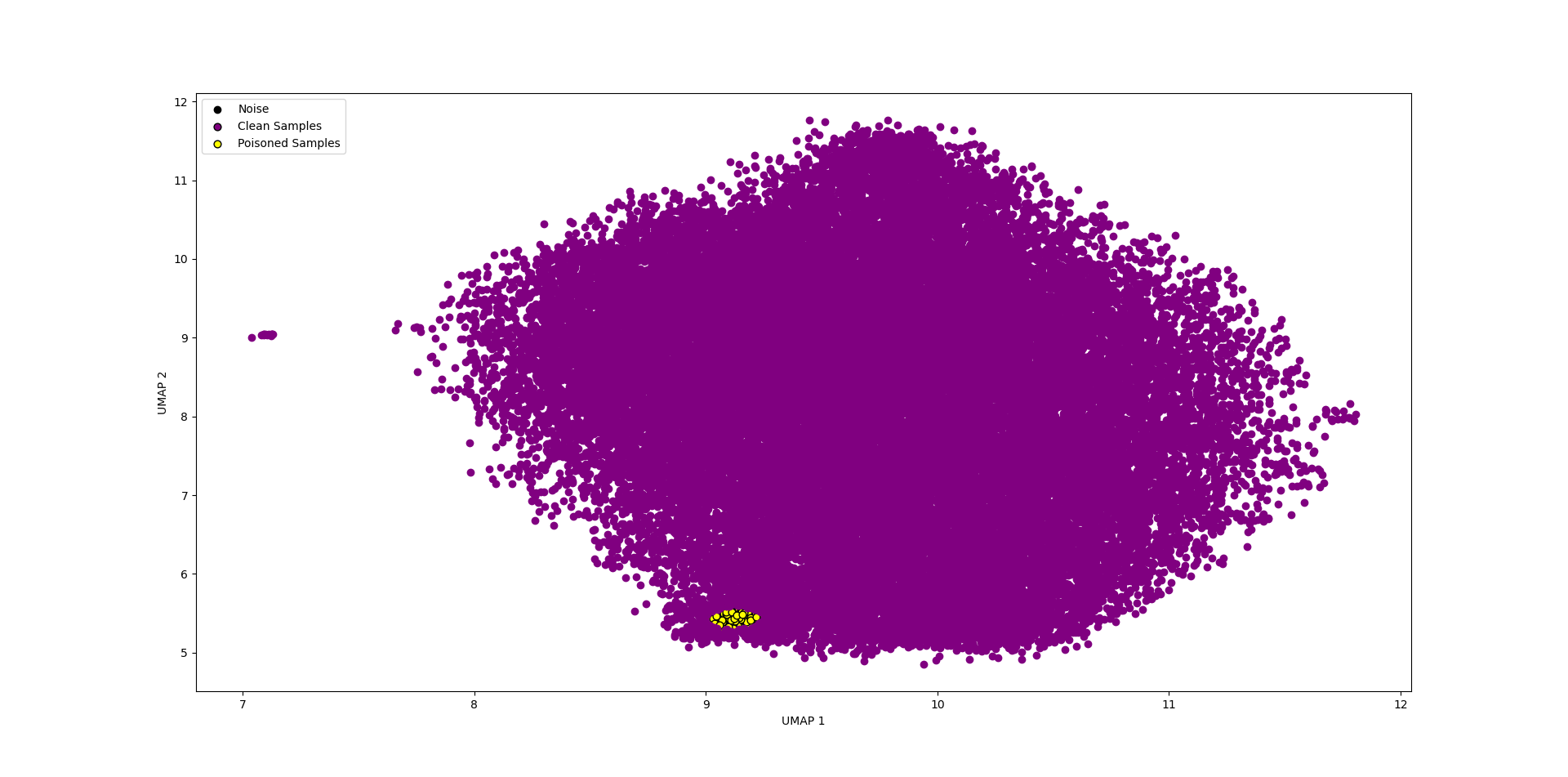} % Adjust width as needed
    \caption{UMAP transformation on the Neurochaos features generated for the poisoned positive class samples of the SST-2 dataset with 5\% poisoning ratio.  NL Hyperparameters Initial Neural Activity = $0.930$, Discrimination threshold = $0.499$, $\epsilon$ = $0.05$.}
    \label{fig:NLhyp_0.05}
\end{figure}

\begin{figure}
    \centering
    \includegraphics[width=\textwidth]{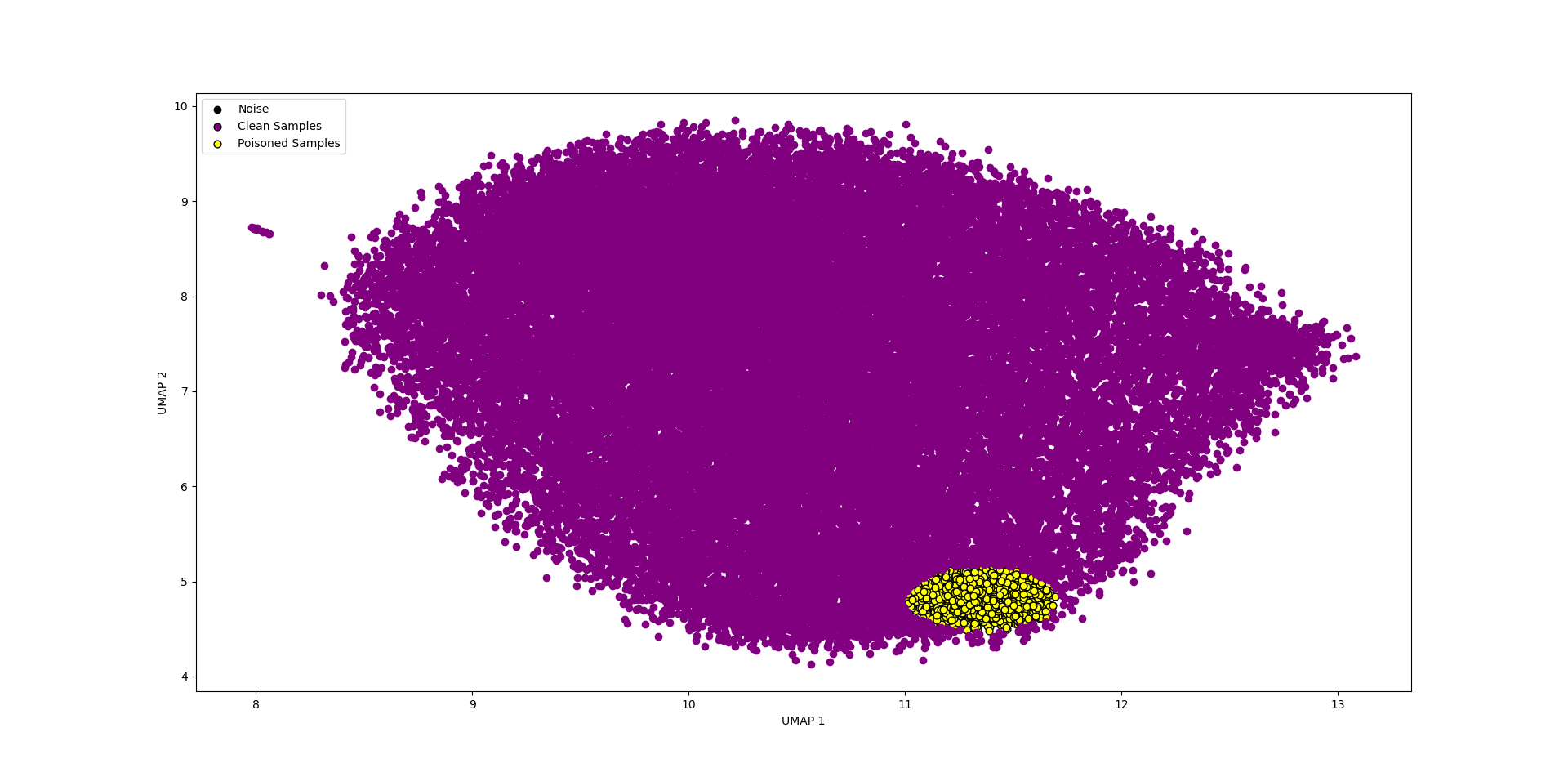} % Adjust width as needed
    \caption{UMAP transformation on the Neurochaos features generated for the poisoned positive class samples of the SST-2 dataset with 5\% poisoning ratio.  NL Hyperparameters Initial Neural Activity = $0.930$, Discrimination threshold = $0.499$, $\epsilon$ = $0.1$.}
    \label{fig:NLhyp_0.1}
\end{figure}

\begin{figure}
    \centering
    \includegraphics[width=\textwidth]{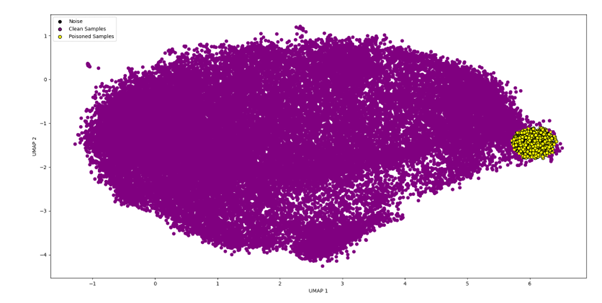} % Adjust width as needed
    \caption{UMAP transformation on the Neurochaos features generated for the poisoned positive class samples of the SST-2 dataset with 5\% poisoning ratio.  NL Hyperparameters Initial Neural Activity = $0.930$, Discrimination threshold = $0.499$, $\epsilon$ = $0.3$.}
    \label{fig:NLhyp_0.2}
\end{figure}

The experimental evaluation of finetuning the NL hyperparameters via the UMAP and DBSCAN transformation steps on the SST – 2 dataset is depicted in Figure~\ref{fig:NLhyp_0.05} through Figure~\ref{fig:NLhyp_0.2}. As observed from Figure~\ref{fig:NLhyp_0.05} through Figure~\ref{fig:NLhyp_0.2}, distinct clusters corresponding to the poisoned and the non-poisoned samples in the poisoned positive class start emerging during the NL hyperparameter tuning stage. The distinct clusters obtained are indicative of presence of backdoors and suggest that the neurochaos features can be used to detect potential backdoors in the training dataset. The NL hyperparameters used for the analysis include the initial neural activity ($q$), discrimination threshold ($b$), and noise intensity threshold ($\epsilon$). The hyperparameter ($\epsilon$) represents the neighborhood used by the GLS neuron to come to a halt or stop firing after starting from an initial neural activity. The best NL hyperparameters $q$, and $b$ were found to be $0.93$ and $0.499$ respectively, whereas $\epsilon$ ranged from $0.3$ to $0.4$ for the NLP datasets. As observed from Figure~\ref{fig:NLhyp_0.05}  through Figure~\ref{fig:NLhyp_0.2} , while holding $q$ and $b$ constant, and by tuning the $\epsilon$ hyperparameter, the UMAP transformation starts detecting poisoned clusters from the neurochaos features.

Upon the computation of the optimal neurochaos features, the corresponding \textit{Precision Matrix Dependency Score (PDS)} is computed for the neurochaos features. This work formulates \textit{Precision Matrix Dependency Score (PDS)} which is computed as the trace or sum of the diagonal elements of the Precision  Matrix. The Precision Matrix, which is mathematically calculated as the inverse of the Covariance Matrix, helps in identifying the conditional dependencies and the conditional variance among various feature variables. While the off-diagonal elements of the Precision Matrix represent conditional dependencies, i.e., any existing correlation between two features after accounting for all other feature variables~\citep{Basor}, the diagonal elements of the precision matrix represent the conditional variance of a feature after accounting for all other variables. The precision matrix has been extensively used in identifying spurious correlations among various feature variables as described in ~\citep{PMbrain}.  The negative off diagonal elements of the precision matrix have been used to identify local group membership in a dataset after accounting for the dominant factors in the feature variables as described in ~\citep{PMfinancialdata}.

However, the utility of the Precision  Matrix as a tool to distinguish poisoned and non poisoned samples in training datasets has never been explored before.  The Precision Matrix Dependency Score $(PDS)$ metric formulated in this work quantifies the conditional variance of neurochaos features for distinct classes in the poisoned training datasets and is depicted in Table~\ref{tab:PDS_SST2} through Table~\ref{tab:PDS_Fakenews}.

\begin{table}[!htb]
\centering
\begin{tabular}{p{1.5cm}|p{1cm}|p{3.5cm}|p{3.5cm}|p{3.5cm}}
Poisoning Ratio & Epsilon & PDS of Poisoned Samples (Positive Class) & PDS of Non-Poisoned Samples (Positive Class) & PDS of Non-Poisoned Samples (Negative Class)\\
\hline
5\%   & 0.3   & 117790   & 52470  & 53278  \\
10\%   & 0.4   & 7224	 & 2317  & 2288  \\
\end{tabular}
\caption{\label{tab:PDS_SST2}SST-2 dataset, Precision Matrix Dependency Score $(PDS)$.}  
\end{table}

\begin{table}[!htb]
\centering
\begin{tabular}{p{1.5cm}|p{1cm}|p{3.5cm}|p{3.5cm}|p{3.5cm}}
Poisoning Ratio & Epsilon & PDS of Poisoned Samples (Positive Class) & PDS of Non-Poisoned Samples (Positive Class) & PDS of Non-Poisoned Samples (Negative Class)\\
\hline
5\%   & 0.3   & 2477268   & 49525  & 51191  \\
10\%   & 0.4   & 3639  & 2183  & 2213  \\
\end{tabular}
\caption{\label{tab:PDS_Jigsaw}Jigsaw Toxicity dataset, Precision Matrix Dependency Score $(PDS)$.}  
\end{table}

\begin{table}[!htb]
\centering
\begin{tabular}{p{1.5cm}|p{1cm}|p{3.5cm}|p{3.5cm}|p{3.5cm}}
Poisoning Ratio & Epsilon & PDS of Poisoned Samples (Positive Class) & PDS of Non-Poisoned Samples (Positive Class) & PDS of Non-Poisoned Samples (Negative Class)\\
\hline
5\%   & 0.4   & 37718 & 34068  & 11667  \\
10\%   & 0.4   & 66094567 & 307042  & 624660 \\
\end{tabular}
\caption{\label{tab:PDS_Fakenews}Fakenews Detection dataset, Precision Matrix Dependency Score $(PDS)$.} 
\end{table}

\subsection*{Complementary Analysis via Shannon Entropy}
\begin{figure}[!htb]
    \centering
    \includegraphics[width=\textwidth]{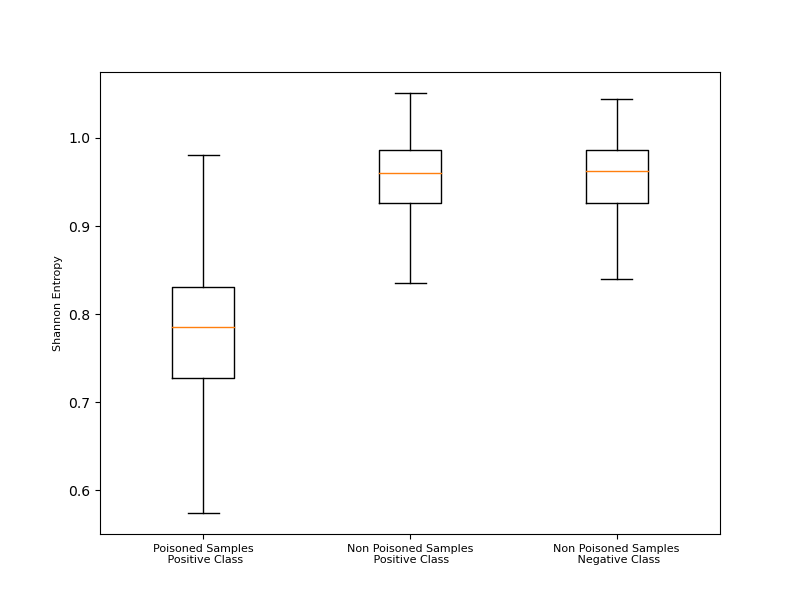} % Adjust width as needed
    \caption{Shannon Entropy computed on the neurochaos features  of the SST-2 dataset, 5\% poisoning ratio.}
    \label{fig:Shannon_SST}
\end{figure}

\begin{figure}[!htb]
    \centering
    \includegraphics[width=\linewidth]{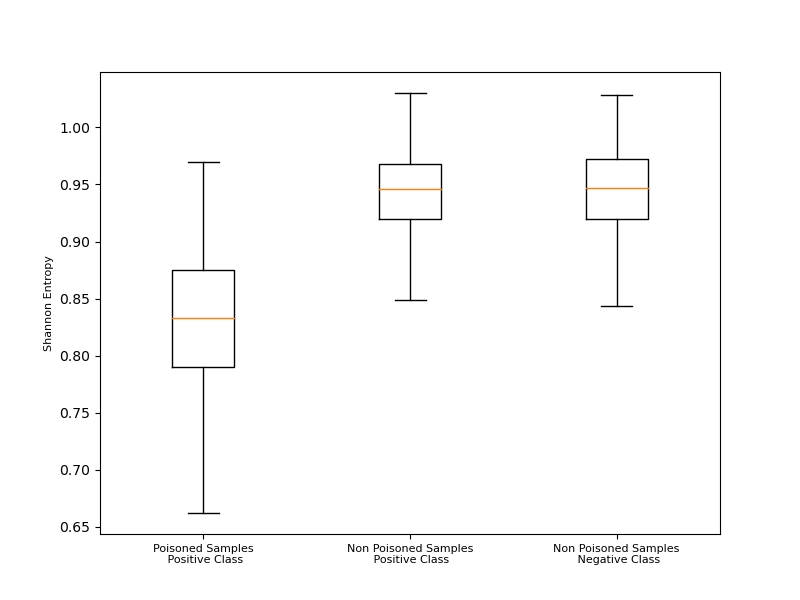} % Adjust width as needed
    \caption{Shannon Entropy computed on the neurochaos features  of the Jigsaw Toxicity dataset, 5\% poisoning ratio.}
    \label{fig:Shannon_Jigsaw}
\end{figure}

\begin{figure}[!htb]
    \centering
    \includegraphics[width=\textwidth]{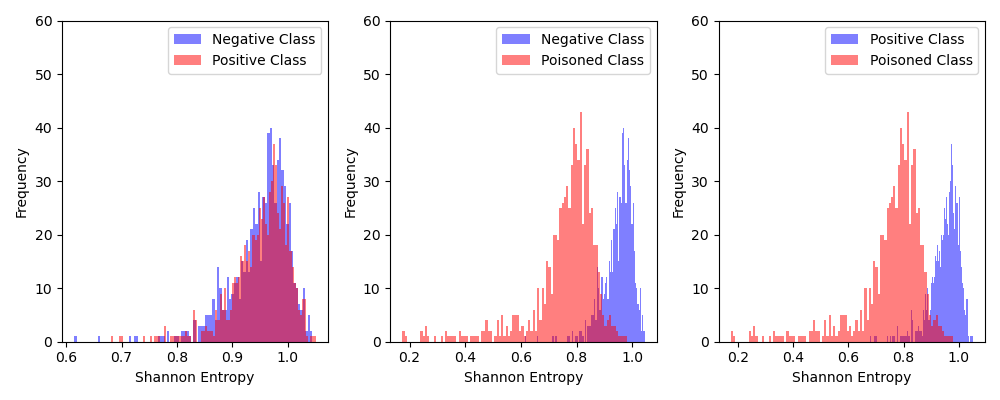} % Adjust width as needed
    \caption{Shannon Entropy distribution computed on the neurochaos features of the SST-2 dataset, 5\% poisoning ratio.}
    \label{fig:Hist_Entropy_SST}
\end{figure}

\begin{figure}[!htb]
    \centering
    \includegraphics[width=\textwidth]{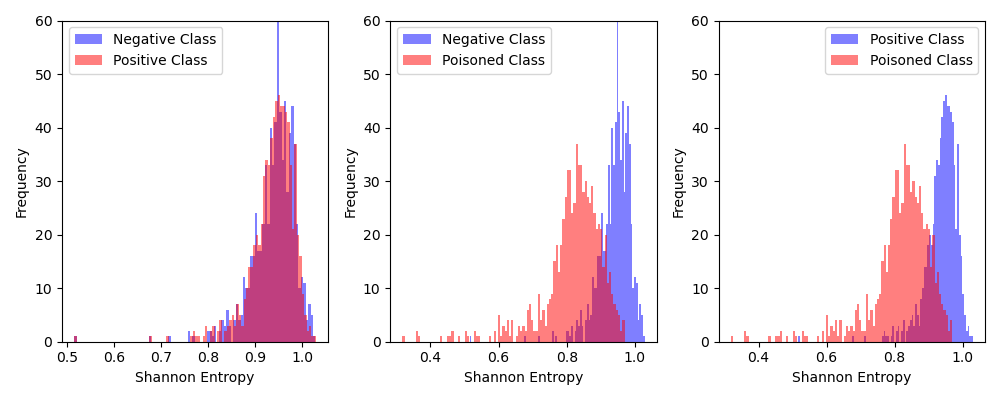} % Adjust width as needed
    \caption{Shannon Entropy distribution computed on the neurochaos features of the Jigsaw Toxicity dataset, 5\% poisoning ratio.}
    \label{fig:Hist_Entropy_Jigsaw}
\end{figure}

The computed $PDS$ values for the poisoned class samples are significantly higher than the non-poisoned class samples, thereby indicating greater predictability and less uncertainty amongst the features. Therefore, a complementary analysis via the Shannon Entropy was performed to validate this finding. Experimental evaluations reveal that the Shannon Entropy computed on the $Energy$ neurochaos feature acts as an effective distinguisher between the poisoned and non-poisoned class. Henceforth the Shannon Entropy analysis was conducted exclusively using the $Energy$ neurochaos feature. The results of the statistical t-tests and Mann Whitney U test performed on the Shannon Entropy for the various classes in the poisoned dataset are depicted in Table~\ref{tab:Shannon_Ttest1} through Table~\ref{tab:Shannon_Ttest5} and Figure~\ref{fig:Shannon_SST} through Figure~\ref{fig:Hist_Entropy_Jigsaw}.

\begin{table}[!htb]
\centering
\begin{tabular}{p{3cm}|p{3cm}|p{3cm}|p{1.5cm}|p{1.5cm}|p{1cm}}
Dataset & Mean Shannon Entropy (Positive Class–Non-Poisoned Samples) & Mean Shannon Entropy (Positive Class-Poisoned Samples)	&t-statistic &p value &Significant difference\\
\hline
SST-2 & $0.95\pm0.05$ & $0.76\pm0.12$	& $40.69$ & $3.96e-217$ & Yes \\
Jigsaw Toxicity & $0.93\pm0.05$ & $0.81\pm0.09$	& $33.82$ & $4.2e-167$ & Yes  \\
Fakenews Detection & $0.46\pm0.15$ & $0.42\pm0.22$	& $13.83$ & $8.01e-41$ & Yes \\
\end{tabular}
\caption{\label{tab:Shannon_Ttest1}t- Test result on on Shannon Entropy of Positive Class (Non-Poisoned Samples) and Positive Class (Poisoned Samples), significance level =0.05, $5\%$ poisoning ratio.}  
\end{table}

\begin{table}[!htb]
\centering
\begin{tabular}{p{3cm}|p{3cm}|p{3cm}|p{1.5cm}|p{1.5cm}|p{1cm}}
Dataset & Mean Shannon Entropy (Positive Class–Non-Poisoned Samples) & Mean Shannon Entropy (Negative Class-Non Poisoned Samples)	&t-statistic & p value &Significant difference\\
\hline
SST-2 & $0.95\pm0.05$ & $0.95\pm0.05$	& $0.08$ & $0.93$ & No \\
Jigsaw Toxicity & $0.93\pm0.05$ & $0.94\pm0.05$	& $-0.78$ & $0.43$ & No  \\
Fakenews Detection & $0.46\pm0.15$ & $0.56\pm0.18$	& $11.56$ & $1.12e-29$ & Yes \\
\end{tabular}
\caption{\label{tab:Shannon_Ttest2}t- Test result on on Shannon Entropy of Positive Class (Non-Poisoned Samples) and Negative Class (Non-Poisoned Samples), significance level =0.05, $5\%$ poisoning ratio.}  
\end{table}

\begin{table}[!htb]
\centering
\begin{tabular}{p{3cm}|p{3cm}|p{3cm}|p{1.5cm}|p{1.5cm}|p{1cm}}
Dataset & Mean Shannon Entropy (Negative Class–Non-Poisoned Samples) & Mean Shannon Entropy (Positive Class–Poisoned Samples)	& t-statistic & p value & Significant difference\\
\hline
SST-2 & $0.95\pm0.05$ & $0.76\pm0.12$	& $40.5$ & $1.12e-216$ & Yes \\
Jigsaw Toxicity & $0.94\pm0.05$ & $0.81\pm0.09$	& $33.29$ & $1.09e-170$ & Yes \\
Fakenews Detection & $0.46\pm0.15$ & $0.42\pm0.22$	& $3.27$ & $0.01$ & Yes  \\
\end{tabular}
\caption{\label{tab:Shannon_Ttest3}t- Test result on Shannon Entropy of Positive Class (Poisoned Samples) and Negative Class (Non-Poisoned Samples), significance level =0.05, $5\%$ poisoning ratio.}  
\end{table}

\begin{table}[!htb]
\centering
\begin{tabular}{p{3cm}|p{3cm}|p{3cm}|p{1.5cm}|p{1.5cm}|p{1cm}}
Dataset & Mean Shannon Entropy (Negative Class – Non-Poisoned Samples) & Mean Shannon Entropy (Positive Class – Poisoned Samples)	& U statistic & p value & Significant difference\\
\hline
SST-2 & $0.95\pm0.05$ & $0.76\pm0.12$	& $573285$ & $4.56e-225$ & Yes \\
Jigsaw Toxicity & $0.94\pm0.05$ & $0.81\pm0.09$	& $548880$ & $1.1e-187$ & Yes \\
Fakenews Detection & $0.56\pm0.18$ & $0.42\pm0.22$	& $315899$ & $0.01$ & Yes  \\
\end{tabular}
\caption{\label{tab:Shannon_Ttest4}Mann-Whitney U Test result on Shannon Entropy of Positive Class (Poisoned Samples) and Negative Class (Non-Poisoned Samples), significance level =0.05, $5\%$ poisoning ratio.}  
\end{table}

\begin{table}[!htb]
\centering
\begin{tabular}{p{3cm}|p{3cm}|p{3cm}|p{1.5cm}|p{1.5cm}|p{1cm}}
Dataset & Mean Shannon Entropy (Negative Class – Non-Poisoned Samples) & Mean Shannon Entropy (Positive Class – Non-Poisoned Samples)& U statistic & p value & Significant difference\\
\hline
SST-2 & $0.95\pm0.05$ & $0.95\pm0.05$	&295814 & $0.91$ & No \\
Jigsaw Toxicity & $0.94\pm0.05$ & $0.93\pm0.05$	& $302903$ & $0.35$ & No \\
Fakenews Detection & $0.56\pm0.18$ & $0.46\pm0.15$	& $200253$ & $1.27e-27$ & Yes  \\
\end{tabular}
\caption{\label{tab:Shannon_Ttest5}Mann-Whitney U Test result on Shannon Entropy of Positive Class (Non-Poisoned Samples) and Negative Class (Non-Poisoned Samples), significance level =0.05, $5\%$ poisoning ratio.}  
\end{table}

The Shannon Entropy distributions for the poisoned class and the non-poisoned class are depicted in Figure~\ref{fig:Hist_Entropy_SST} and Figure~\ref{fig:Hist_Entropy_Jigsaw}.

An additional measure of spread or dispersion $D_p$ of neurochaos features across dimensions was computed by applying the below transformation to  all the normalized  neurochaos features from each distinct class. The neurochaos features studied for this analysis include the $Energy$, $Entropy$, $Firing Time$ and $Firing Rate$. 

\begin{equation}
    D_p =  -\sum_{i=1}^{N} f{p_i} \log f{p_i},
\end{equation}
where $f{p_i}$ represents the normalized neurochaos feature vectors corresponding to the poisoned class, and $N$ represents the total number of features in the poisoned class.  

\begin{figure}[!htb]
    \centering
    \includegraphics[width=\textwidth]{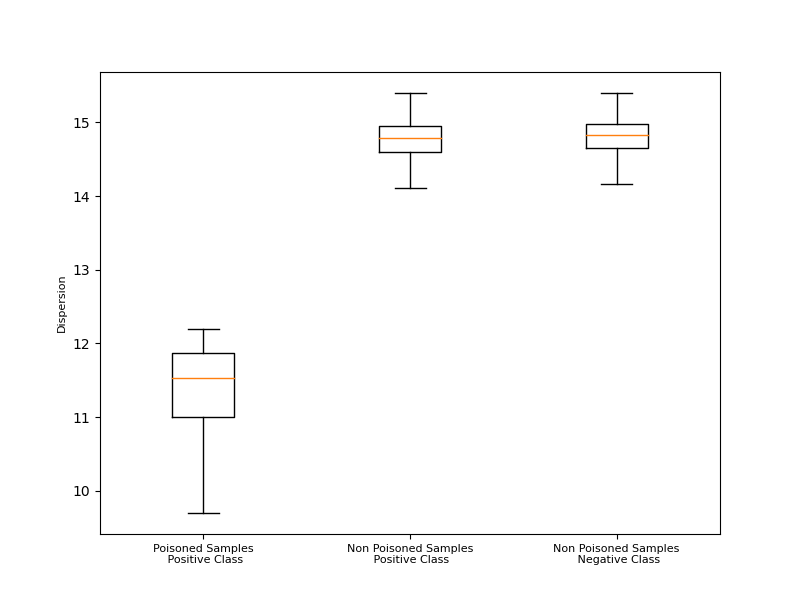} % Adjust width as needed
    \caption{Dispersion Measure computed on the neurochaos features  of the Jigsaw Toxicity dataset, 10\% poisoning ratio.}
    \label{fig:Dispersion}
\end{figure}

The measure of dispersion is computed across the three classes and the results are depicted in Figure~\ref{fig:Dispersion}. As observed from Figure~\ref{fig:Dispersion}. 

\section*{Discussion}

\subsection*{Analysis and Interpretation of PDS}
As observed from Table~\ref{tab:PDS_SST2} through Table ~\ref{tab:PDS_Fakenews}, the $PDS$ that is calculated for positive class poisoned samples is significantly higher than both the positive class and negative class non-poisoned samples for various poisoning ratios. This observation holds good for all of the NLP datasets. The $PDS$ effectively quantifies the conditional variance of the neurochaos features for each distinct class in the poisoned training dataset. Given this fact, the extremely high values of $PDS$ observed for the poisoned class samples in comparison to the non-poisoned class samples are indicative of very high precision for the neurochaos features from the poisoned class. High Precision in the diagonal elements of the Precision Matrix in turn implies that there is very low conditional variance for the corresponding features. This finding suggests that in the Precision Matrix of the poisoned class, the conditional variance of the diagonal elements has very little variability when all other features are accounted for.  Hence the neurochaos features for the poisoned class samples have less variability and are more predictable once all the other features are known/ conditioned upon. This implies that the neurochaos features corresponding to the poisoned class samples have high dependency amongst them, when compared to the non-poisoned class samples.

\subsection*{Interpretation of Shannon Entropy}

As observed from Figure~\ref{fig:Shannon_SST} and Figure~\ref{fig:Shannon_Jigsaw}, the mean Shannon Entropy of the poisoned class neurochaos features is comparatively less than the mean Shannon Entropy of the non-poisoned class neurochaos features across datasets indicating less uncertainty for the poisoned class samples. Therefore, the mean Shannon Entropy can be employed to validate the $PDS$ measure of dependency and low conditional variability observed for the poisoned class samples. Thus, the neurochaos features obtained via the Chaos based – PDS approach can be potentially used to detect backdoor triggers in the training datasets. 

Additionally, the Shannon Entropy distribution plots in Figure~\ref{fig:Hist_Entropy_SST} and Figure~\ref{fig:Hist_Entropy_Jigsaw} for the poisoned class and the non-poisoned class support the finding that the Shannon Entropy computed on the neurochaos features can be utilized to separate the poisoned and the non-poisoned classes in the training datasets. 

The measure of dispersion is computed across the three classes and the results are depicted in Figure~\ref{fig:Dispersion}. As observed from Figure~\ref{fig:Dispersion}, the dispersion measure which indicates the variability in the neurochaos features aligns well with the computed Shannon Entropy measures depicted in Figure~\ref{fig:Shannon_SST} and Figure~\ref{fig:Shannon_Jigsaw}.

\section*{Conclusions}
The assumptions of the threat models used by existing defense mechanisms include the defender's reliance on trusted datasets, and/or model training to detect backdoor triggers in the training dataset.  Given these limitations, the current work proposes a novel integrated approach based on chaos theory and manifold learning to distinguish poisoned samples from the non-poisoned samples of the training dataset. In the chaos based approach for backdoor detection, a novel backdoor defense mechanism is proposed to address and overcome the existing limitations inherent in the existing backdoor defense measures. It should be noted that this defense mechanism operates without the necessity to train an ML model or requiring access to a training data set.To this end, a novel \textit {Precision Matrix Dependency Score (PDS)} metric based on the conditional variance of the neurochaos features was formulated. 

The experimental evaluation of the proposed $PDS$ metric demonstrates the efficacy of the proposed measures in detecting static backdoor triggers across diverse datasets from the NLP domain. The validity of the $PDS$ metric was further verified using the Shannon Entropy measure. Future work will include extending the Chaos based approach to other diverse forms of backdoor triggers.

%\bibliography{references}

\end{document}